\documentclass[twocolumn,10pt]{IEEEtran}

\usepackage{calc}

\usepackage{xcolor,multirow,gensymb}
  \usepackage{cite}
\usepackage{graphicx,subfigure,epsfig}
\usepackage{psfrag}
\usepackage{amsmath,amssymb}
\usepackage{color}
\usepackage{array}
\usepackage{nicefrac}
\usepackage{amsmath,amssymb}
\usepackage{xifthen}
\usepackage{mathtools}
\usepackage{enumerate}
\usepackage{microtype}
\usepackage{xspace}
\usepackage{bm}
\usepackage[T1]{fontenc}
\usepackage{fancyhdr}
\usepackage{lastpage}
\usepackage{rotating}
\usepackage{tabulary}

\newcommand{\vect}[1]{{\lowercase{\mbs{#1}}}}

\newcommand{\mbs}[1]{\bm{#1}}
\newcommand{\mat}[1]{{\uppercase{\mbs{#1}}}}
\newcommand{\Id}{\mat{\mathrm{I}}}

\newcommand{\T}{{\scriptscriptstyle\mathsf{T}}}
\renewcommand{\H}{{\scriptscriptstyle\mathsf{H}}}

\renewcommand{\Re}[1][]{\ifthenelse{\isempty{#1}}{\operatorname{Re}}{\operatorname{Re}\left(#1\right)}}
\renewcommand{\Im}[1][]{\ifthenelse{\isempty{#1}}{\operatorname{Im}}{\operatorname{Im}\left(#1\right)}}
\usepackage{float}

\newfloat{longequation}{b}{ext}

\newcommand{\ev}{\vect{e}}

\newcommand{\deltav}{\hbox{\boldmath$\delta$}}

\def\bD{{\mathbf{D}}}

\def\bF{{\mathbf{F}}}
\def\bG{{\mathbf{G}}}

\def\bK{{\mathbf{K}}}

\def\bR{{\mathbf{R}}}
\def\bS{{\mathbf{S}}}
\def\bT{{\mathbf{T}}}

\def\bW{{\mathbf{W}}}
\def\bX{{\mathbf{X}}}

\def\bZ{{\mathbf{Z}}}
\def\bSigma{{\mathbf{\Sigma}}}

\newcommand{\cC}{{\cal C}}

\newcommand{\cN}{{\cal N}}

\def\bb{{\mathbf{b}}}

\def\bee{{\mathbf{e}}}
\def\bff{{\mathbf{f}}}
\def\bg{{\mathbf{g}}}
\def\bh{{\mathbf{h}}}

\def\bw{{\mathbf{w}}}
\def\bx{{\mathbf{x}}}
\def\by{{\mathbf{y}}}
\def\bz{{\mathbf{z}}}
\def\b0{{\mathbf{0}}}

\def\bTheta{{\boldsymbol{\Theta}}}

\def\bbC{{\mathbb{C}}}

\def\bpsi{{\boldsymbol{\psi}}}

\newcommand{\EE}{\mathbb{E}}

\newcommand{\CN}[1][]{\ifthenelse{\isempty{#1}}{\mathcal{N}_{\mathbb{C}}}{\mathcal{N}_{\mathbb{C}}\left(#1\right)}}

\renewcommand{\P}[1][]{\ifthenelse{\isempty{#1}}{\mathbb{P}}{\mathbb{P}\left(#1\right)}}
\newcommand{\E}[1][]{\ifthenelse{\isempty{#1}}{\mathbb{E}}{\mathbb{E}\left(#1\right)}}
\renewcommand{\det}[1][]{\ifthenelse{\isempty{#1}}{\text{det}}{\text{det}\left(#1\right)}}
\newcommand{\trace}[1][]{\ifthenelse{\isempty{#1}}{\text{tr}}{\text{tr}\left(#1\right)}}
\newcommand{\rank}[1][]{\ifthenelse{\isempty{#1}}{\text{rank}}{\text{rank}\left(#1\right)}}
\newcommand{\diag}[1][]{\ifthenelse{\isempty{#1}}{\text{diag}}{\text{diag}\left(#1\right)}}
\def\nn{\nonumber}

\IEEEoverridecommandlockouts

\newtheorem{proposition}{Proposition}
\newtheorem{remark}{Remark}
\newtheorem{lemma}{Lemma}

\def\bPhi{{\boldsymbol{\Phi}}}

\newcommand{\tr}{\mathop{\mathrm{tr}}\nolimits}
\newtheorem{Theorem}{Theorem}

\newcommand{\al}{\alpha}

\newcounter{enumi_saved}
\usepackage{answers}
\Newassociation{solution}{Solution}{solutionfile}

\AtBeginDocument{\Opensolutionfile{solutionfile}[\jobname]}
\AtEndDocument{\Closesolutionfile{solutionfile}\clearpage
}

\DeclareSymbolFont{matha}{OML}{txmi}{m}{it}
\DeclareMathSymbol{\varv}{\mathord}{matha}{118}
\usepackage{amsmath}
\newcommand{\overbar}[1]{\mkern 1.5mu\overline{\mkern-1.5mu#1\mkern-1.5mu}\mkern 1.5mu}
\renewcommand{\baselinestretch}{1.02}

\begin{document}
\title{Mitigation of Phase Noise in Massive MIMO Systems: A Rate-Splitting Approach}
\author{Anastasios Papazafeiropoulos, Bruno Clerckx, and Tharm Ratnarajah   \vspace{2mm} \\
\thanks{A. Papazafeiropoulos and T. Ratnarajah are  with the  Institute for Digital Communications (IDCOM), University of Edinburgh, Edinburgh, EH9 3JL, U.K., (email: {a.papazafeiropoulos, t.ratnarajah}@ed.ac.uk). B. Clerckx is with the Communications and Signal Processing group in the Department of Electrical and Electronic
Engineering, Imperial College London, SW7 2AZ U.K., (email: b.clerckx@imperial.ac.uk.)}
\thanks{This work was supported by the U.K. Engineering and Physical Sciences Research Council (EPSRC) under grants EP/N014073/1 and EP/N015312/1.}}
\maketitle
\begin{abstract}
This work encompasses Rate-Splitting (RS), providing significant benefits in multi-user settings in the context of huge degrees of freedom promised by massive  Multiple-Input Multiple-Output (MIMO). However, the requirement of massive MIMO for cost-efficient implementation makes them more prone to hardware imperfections such as phase noise (PN). As a result, we focus on a realistic broadcast channel with a large number of antennas and hampered by the unavoidable PN. Moreover, we employ the  RS transmission strategy, and we show its robustness against PN, since the sum-rate does not saturate at  high  signal-to-noise ratio (SNR). Although, the analytical results are obtained by means of the deterministic equivalent analysis, they coincide with simulation results even for finite system dimensions. 
\end{abstract}

\section{Introduction}
Massive Multiple-Input Multiple-Output (MIMO), known also as large MIMO, is one of the promising technologies for Fifth Generation, (5G) networks~\cite{METIS}. The key concept takes into account the law of large numbers. Specifically, a base station (BS) with a large number of antennas enables fast fading, intra-cell interference, and additive Gaussian noise to averaged out as the number of antennas tends to infinity~\cite{Rusek2013,Hoydis2013,Papazafeiropoulos2015a}. 

Inevitable phase noise (PN) occurs in communication systems even after applying calibration and compensation techniques~\cite{Studer2010}. It is a fundamental bottleneck of wireless communications that cannot be estimated with infinite precision. PN includes phase drifts from the Local Oscillators (LOs) that present a multiplicative nature with regards to the channel vector. Note that phase drifts accumulate within the channel coherence time.    PN contributes to inaccurate Channel State Information at the Transmitter (CSIT), and degrades further the spectral efficiency. Moreover, its effect becomes more severe in massive MIMO systems involving a large number of antennas. In fact, the more cost-efficient massive MIMO are, the more prone to hardware impairments, such as PN, are. Actually, PN has been considered in some works such as~\cite{Pitarokoilis2015,Krishnan2015,Papazafeiropoulos2016} to assess the realistic performance of communication systems, but none of them has accounted for its mitigation.  Unfortunately, the majority of massive MIMO literature has assumed perfect hardware, despite the existence of PN. It is conjectured that if we follow the same path, the gap between theory and practice will increase, and misleading conclusions will be made during the design and evaluation of 5G systems. 

This work tackles the challenge of mitigating PN by leveraging the Rate-Splitting (RS) approach. According to RS, we can split one's User-Element (UE) message into a common part and a private part. RS outperforms conventional broadcasting because it does not experience any ceiling effect.  Henceforth, we denote by NoRS all the  conventional techniques to contrast with the RS technique. We aim at showing the robustness of RS in massive MIMO systems by a deterministic equivalent (DE) analysis, when PN is accounted. In particular, this work shows the robustness of the RS method in realistic Time-Division-Duplex (TDD)-based massive MIMO with PN and imperfect CSIT. Note that both pilot contamination and PN contribute to imperfect CSIT.

The   remainder of this paper is structured as follows.  Section~\ref{SystemModel} presents the system  model. Moreover, we present the PN and the RS approach.  Next, in Section~\ref{training}, we provide the uplink training phase with  PN, while Section~\ref{Downlink} shows the corresponding downlink transmission.  Section~\ref{Deterministic} exposes the design of the  precoder of the common message, and mainly, the  achievable rates in the presence of PN in terms  of a DE analysis. The numerical results are placed in Section~\ref{NumericalResults}, while Section~\ref{Conclusions} summarizes the paper.

\textit{Notation:} Vectors and matrices are denoted by boldface lower and upper case symbols. $(\cdot)^\T$, $(\cdot)^*$,  $(\cdot)^\H$, and $\tr\!\left( {\cdot} \right)$ represent the transpose, conjugate, Hermitian  transpose, and trace operators, respectively. The expectation  operator is denoted by $\EE\left[\cdot\right]$. The $\mathrm{diag}\{\cdot\}$ operator generates a diagonal matrix from a given vector, and the symbol $\triangleq$ declares definition. The notations $\mathcal{C}^{M \times 1}$ and $\mathcal{C}^{M\times N}$ refer to complex $M$-dimensional vectors and  $M\times N$ matrices, respectively. Finally, $\bb \sim \cC\cN{(\b0,\mathbf{\Sigma})}$ and $\bb \sim \cN{(\b0,\mathbf{\Sigma})}$ denote a circularly symmetric complex Gaussian variable with zero-mean and covariance matrix $\mathbf{\Sigma}$ and the corresponding real Gaussian variable, respectively.



\section{System Model}\label{SystemModel} 
We consider a  Broadcast (BC) channel, where the BS has $M$ antennas and serves simultaneously $K$ single-antenna UEs. Moreover, we assume that the channel $\bh_{k}\triangleq \left[h_{k}^{1}, \ldots, h_{k}^{M}  \right]\in \bbC^{M\times 1}$ between the BS and UE $k$  is a frequency-flat channel, expressed by
\begin{align}
\bh_{k} = \bR^{1/2}_{k}\bw_{k},
\end{align}
where $\bR_{k}\! =\!\mathbb{E}\!\left[ \bh_{k} \bh_{k}^{\H}\right]\!\in\! \bbC^{M \times M }$ is a deterministic Hermitian-symmetric positive-definite matrix representing versatile effects such as the path loss to each antenna.  Note that $\bw_{k} \in \bbC^{M \times 1}$  is an uncorrelated fast-fading Gaussian channel vector drawn as  $\bw_{k} \sim \cC\cN(\b0,\Id_{M})$. Hence, we have that $\bh_{k} \sim \cC\cN \left( \b0,\bR_{k} \right)$.

\subsection{Phase Noise}\label{PN1} 
The PN expresses the distortion in the phase due to the random phase drift  in the signal coming from the LOs of the BS and UE $k$, and it is induced during the up-conversion of the baseband signal to passband and vice-versa. Mathematically, it is described by a discrete-time  independent Wiener process~\cite{Demir2000,Pitarokoilis2015}. Specifically,  the PNs at the  LOs of the $m$th antenna of the BS and $k$th UE are modeled as 
\begin{align}
 \phi_{m,n}&=\phi_{m,n-1}+\delta^{\phi_{m}}_{n}\label{phaseNoiseBS}\\
 \varphi_{k,n}&=\varphi_{k,n-1}+\delta^{\varphi_{k}}_{n},\label{phaseNoiseuser}
\end{align}
where $\delta^{\phi_{m}}_{n}\sim \cN(0,\sigma_{\phi_{m}}^{2}) $ and $\delta^{\varphi_{k}}_{n}\sim \cN(0,\sigma_{\varphi_{k}}^{2})$. Note that $\sigma_{i}^{2}=4\pi^{2}f_{\mathrm{c}} c_{i}T_{\mathrm{s}}$, $i=\phi_{m}, \varphi_{k}$ describes the PN increment variance with $T_{\mathrm{s}}$, $c_{{i}}$, and $f_{\mathrm{c}}$ being the  symbol interval, a constant dependent on the oscillator, and the carrier frequency, respectively.

We assume that the  PN processes are considered as mutually independent, if each antenna has its own oscillator, i.e., a  Separate Local Oscillator (SLO) at each antenna.  In the case that we have just one Common LO (CLO) connected to all BS antennas, there is only one PN process $\phi_{n}$. In our analysis, we focus  on both SLOs and CLO scenarios, but in all cases, we assume  i.i.d. PN statistics across different antennas and UEs, i.e., $\sigma_{\phi_{m}}^{2}=\sigma_{\phi}^{2}$ and $ \sigma_{\varphi_{k}}^{2}=\sigma_{\varphi}^{2}~,\forall~m,~k$.

Actually, the PN is expressed as a multiplicative factor to the channel vector as 
\begin{align}
\tilde{\bg}_{k,n}=\bTheta_{k,n}\bh_{k}\label{PhaseNoise},
\end{align}
where $\bTheta_{k,n}\!\triangleq\!\mathrm{diag}\!\left\{ e^{j \theta_{k,n}^{(1)}}, \ldots, e^{j \theta_{k,n}^{(M)}} \!\right\}=e^{j \varphi_{k,n}}\bPhi_{n}\in \mathbb{C}^{M\times M}$ is the total PN with $\bPhi_{n}\!\triangleq\!\mathrm{diag}\!\left\{\! e^{j \phi_{1,n}}, \ldots, e^{j \phi_{M,n}} \!\right\}$ being the PN sample matrix  at time $n$ because of the imperfections in the LOs of the BS, while, $e^{j \varphi_{k,n}}$ is the PN induced by UE $k$. Basically, $\tilde{\bg}_{k,n}$ represents the effective channel vector at time $n$. Clearly, the effective channel, given by \eqref{PhaseNoise}, depends on the time slot of symbol $n$ due to the time-dependence coming from the PN. 

\subsection{RS Approach}
RS is a very promising method in multi-user transmissions with imperfect CSIT, since it achieves unsaturated sum-rate with increasing SNR despite the presence of imperfect CSIT~\cite{Hao2015,Dai2016,Clerckx2016}. 

According to this method,  the message intended for UE $k$ is split into two parts, namely, a common and a private part. The  common part, drawn from a public codebook,  should be decoded by all UEs with zero error probability. The private part is to be decoded only by UE $k$. Note that the  messages intended for the other UEs  consist of a private part only. In mathematical terms, we have
\begin{align}
\bx=\underbrace{\sqrt{\rho_{\mathrm{c}}}\bff_{c} s_{c}}_{\mathrm{common ~part}}+\underbrace{\sum_{k=1}^{K}\sqrt{\lambda \rho_{\mathrm{k}}}\bff_{k} s_{k}}_{private~part},\label{RStransmit}
\end{align}
where $s_{c}$ is the common message and $s_{k}$ is the private message of UE $k$, while $\bff_{c}$ denotes the  precoding vector of the common
message  with unit norm and $\bff_{k}$ is the linear precoder corresponding to UE $k$. Note that $\lambda$ is the normalization parameter regarding the precoder given by
\begin{align}
\lambda=\frac{K}{\EE \left[ \tr\bF^\H\bF \right]}. \label{eq:lamda} 
\end{align} According to the decoding procedure,  the common message is decoded by each UE, while all private messages are treated as noise. Next, each UE subtracts the contribution of the common message in the received signal and is able to decode its own private message.

\section{Uplink Pilot Training Phase with PN}\label{training} 
By assuming TDD, we consider coherence blocks with duration of $T$ channel uses. Each block is split into   $\tau \ge K$ uplink pilot symbols and $T-\tau$ downlink data symbols. The CSI is  acquired during the uplink training phase, while we exploit channel reciprocity for the downlink channel. During this phase, we assign a pilot sequence of $\tau$ symtbols  to  UE $k$, i.e., $\bm \omega_{k}\triangleq \left[\omega_{k,1},\ldots,\omega_{k,\tau} \right]^{\T}\in \bbC^{\tau \times 1}$ with $\rho_{up}^{\mathrm{UE}}=\left[|\omega_{k,n}|^{2} \right],\forall k,n$. Note that the sequences among different UEs are mutually orthogonal.

The received uplink vector at the BS at time $n \in \left[0, \tau \right]$ $\by^{\mathrm{tr}}_{n} \in \bbC^{M \times 1}$, accounting for the PN, is given by 
\begin{align}
&\!\!\!\by^{\mathrm{tr}}_{n}\!=\!\sum_{k=1}^{K}\tilde{\bg}_{k,n}\omega_{k,n}+\bz_{n}^{\mathrm{BS}} \label{BasicSystemModel},
\end{align}
where  $\bz_{n}^{\mathrm{BS}}\sim \mathcal{CN}\left( \b0,\sigma_{\mathrm{BS}}^{2} \Id_{M} \right)$ is the Additive White Gaussian Noise (AWGN) at UE $k$. As mentioned, $\bh_{k}$ is assumed to be constant during the coherence time $T$, while it  changes independently afterwards.

Concatenation of all the received signal vectors during the training phase results in a new vector $\bpsi \triangleq \left[{\by^{\mathrm{tr}}_{0}}^{\T}, \ldots, {\by^{\mathrm{tr}}_{\tau}}^{\T} \right]^{\T} \in \bbC^{\tau M\times 1} $. Simlar to~\cite{Bjornson2015}, the  Linear Minimum Mean-Square Error (LMMSE) estimate of the channel of UE $k$ during the training phase is given by
\begin{align}
\hat{\bg}_{k,n}&=\EE\left[\tilde{\bg}_{k,n}\bm \psi^{\H} \right]\left( \EE\left[\bm \psi \bm \psi^{\H} \right]  \right)^{-1}\bm \psi\nn\\
&= \left( \bm \omega_{k}^{\H}\bm \Delta_{k}^{\mathrm{\tr}}\otimes \bR_{k}\right)\bm \Sigma^{-1}\bm \psi\label{EstimatedChannel},
\end{align}
where 
\begin{align}
 \bm \Delta_{k}^{\mathrm{\tr}}&\triangleq\mathrm{diag}\!\left\{ e^{-\frac{\sigma_{\varphi}^{2}+\sigma_{\phi}^{2}}{2}}, \ldots, e^{-\frac{\sigma_{\varphi}^{2}+\sigma_{\phi}^{2}}{2}\tau} \!\right\}\\
 \bm \Sigma&\triangleq \sum_{j=1}^{K}\bX_{j}\otimes\bR_{j}+\sigma_{\mathrm{BS}}^{2}\Id_{\tau M},\\
 \bD_{|\bm \omega_{j}|^{2}}&\triangleq \mathrm{diag}\left( |\omega_{j,1}|^{2},\ldots,|\omega_{j,\tau}|^{2} \right),\\
 \left[ {\bX}_{j}\right] _{u,v}&\triangleq \omega_{j,u}\omega_{j,v}^{*}\rho_{up}^{\mathrm{UE}}e^{-\frac{\sigma_{\varphi}^{2}+\sigma_{\phi}^{2}}{2}|u-v|}.
\end{align}
\begin{proof}
The proof is ommited for the sake of limited space.
\end{proof}

The LMMSE estimation enables us to write the current channel at the end of the training phase as
\begin{align}
\tilde{\bg}_{k,\tau}=\hat{\bg}_{k,\tau}+\bee_{k,\tau},\label{current}                                    
\end{align}
where $\bee_{k,\tau}$ is the Gaussian distributed zero-mean estimation error vector with covariance given by $\tilde{\bR}_{k}=\bR_{k}-\hat{\bR}_{k}$.

We have $\hat{\bg}_{k}\!\sim\! \cC\cN \left( \b0,\hat{\bR}_{k} \right)$ with $\hat{\bR}_{k}=\left( \bm \omega_{k}^{\H}\bm \Delta_{k}^{\mathrm{\tr}}\otimes \bR_{k}\right)\bm \Sigma^{-1}$ $ \left( {\bm \Delta_{k}^{\mathrm{\tr}}}^{\H } \bm \omega_{k} \otimes  \bR_{k} \right)$.

The dependence of the estimated channel on time $n$ necessitates a continuous computation of the applied precoder in the downlink at every  symbol interval, which is computationally prohibitive due to its complexity. Therefore, we assume  that the precoder is designed by means of the channel estimate once during the training phase
and then is applied for the whole duration of the downlink transmission phase. For example, if the channel is estimated at $n_{0}=\tau$, the applied precoder is denoted by $\bff_{k  }\triangleq\bff_{k,n_{0}+1}$.
\section{Downlink Transmission under PN}\label{Downlink}
Based on channel reciprocity, the received signal by UE $k$ during the transmission phase $n \in \left[ \tau+1, T\right] $ is given by
\begin{align}
y_{k,n}={\bh}_{k}^{\H}{\bTheta}_{k,n}^{*} \bx+ z_{k,n}^{\mathrm{UE}}, \label{BasicSystemModelDownlink}
\end{align}
where $z_{k,n}^{\mathrm{UE}}\sim\left( 0,\sigma_{\mathrm{UE}}^{2} \right)$ is the Additive White Gaussian Noise (AWGN) at UE $k$.

During the downlink transmission phase described by~\eqref{BasicSystemModelDownlink}, we set $
 {\bh}_{k}^{\H}{\bTheta}_{k,\tau}^{*}= \tilde{\bg}^{\H}_{k,\tau}$. If we solve with respect to ${\bh}_{k}^{\H}$ and make the necessary substitution, we result in 
\begin{align}
 {\bh}_{k}^{\H}{\bTheta}_{k,n}^{*}= \tilde{\bg}_{k,\tau}^{\H}\widetilde{\bTheta}_{k,n}.
\end{align}
where $\widetilde{\bTheta}_{k,n}\!\triangleq\!\mathrm{diag}\!\left\{ e^{-j \left( \theta_{k,n}^{(1)}-\theta_{k,\tau}^{(1)} \right)}, \ldots, e^{-j \left( \theta_{k,n}^{(M)}-\theta_{k,\tau}^{(M)} \right)}\!\right\}$. Consequently, if we set $\bg_{k,n}= \widetilde{\bTheta}_{k,n}^{*}\tilde{\bg}_{k,\tau}$~\eqref{BasicSystemModelDownlink} becomes
\begin{align}
y_{k,n}={\bg}_{k,n}^{\H} \bx + z_{k,n}^{\mathrm{UE}}. \label{BasicSystemModelDownlink1}
\end{align}

The trace ${T}_{\mathrm{PN}}$ of PN is given by
\begin{align}
{T}_{\mathrm{PN}}&=\tr \widetilde\bTheta_{k,n}\nn\\
&=\sum_{l=1}^{M}e^{-j \left(  \theta_{k,n}^{(l)}-\theta_{k,\tau}^{(l)} \right)}\label{PNtrace}.
\end{align}
From~\eqref{PNtrace}, we get the following useful lemma. 
\begin{lemma}\label{SetupLOs}
For CLO and SLOSs, we have 
\begin{align}
 \frac{1}{M}T_{\mathrm{PN}} \xrightarrow[ M \rightarrow \infty]{} \begin{cases}
                            e^{-j \left(\delta^{\phi}+\delta^{\varphi}\right)}~~~&\mathrm{CLO~setup}\\
                             e^{- \frac{\delta_{\phi}^{2}}{2}n-j \delta^{\varphi}}~~~&\mathrm{SLOs~setup}.                         \end{cases}
\end{align}
\end{lemma}
\proof The proof is straightforward by means of the application of the law of large numbers.
\endproof
\begin{remark}
This lemma describes the effect of PN from both BS and UE LOs.
\end{remark}



\subsection{SINR with PN and NoRS (Conventional Transmission)} \label{RS} 
The SINR  of UE $k$, assuming equal power allocation, is expressed by means of~\eqref{BasicSystemModelDownlink1}  as 
 \begin{align}
   \mathrm{SINR}_{k,n}^{\mathrm{NoRS}}=\frac{\frac{\rho_{k}}{K}{\lambda}|\bg_{k,n}^{\H}\bff_{k}|^2}{{\lambda}\sum_{j\ne k}^{K}\frac{\rho_{j}}{K}|\bg_{k,n}^{\H}\bff_{j}|^2 +\sigma_{\mathrm{UE}}^{2}}.\label{nors} 
\end{align}

Note that we  treat the multi-user interference as independent Gaussian noise (the worst-case assumption) for the calculation of the mutual information~\cite[Lemma 1]{Bjornson2015}.

The  mutual information between the received signal and the transmitted symbols is lower bounded by 
\begin{align}
 \mathrm{R}^{\mathrm{NoRS}}&=\sum_{k=1}^{K}\mathrm{R}_{k}^{\mathrm{NoRS}}\nn\\
 &=\frac{1}{T_{c}}\sum_{k=1}^{K}\sum_{n=1}^{T_{c}-\tau}\mathrm{R}_{k,n}^{\mathrm{NoRS}},\label{rate3} 
\end{align}
where $\mathrm{R}_{k,n}^{\mathrm{NoRS}}=\log_{2}\left( 1+ \mathrm{SINR}_{k,n}^{\mathrm{NoRS}} \right)$. In particular,  we compute the achievable rate of each UE for each time instance of the data transmission phase as in~\cite{Bjornson2015,Pitarokoilis2015}.

\subsection{SINR with PN under RS}\label{RS} 
We allocate $\rho_{\mathrm{c}}=\rho\left( 1-t \right)$ to the common message and  $\rho_{\mathrm{k}}=\rho t/K$ to the private message of each UE, where $t \in \left( 0,1 \right]$. The role of $t$ is to adjust the  fraction of the total power spent for the transmission of the private messages.

The SINRs of both common and private messages  are given  by
\begin{align}
   \mathrm{SINR}_{k,n}^{\mathrm{c}}&=\frac{{\rho_{\mathrm{c}}}{\lambda}|\bg_{k,n}^{\H} \bff_{c}|^2}{\lambda\sum_{j=1}^{K}{\frac{\rho_{j}}{K}|\bg_{k,n}^{\H} \bff_{j}|^2+ \sigma_{\mathrm{UE}}^{2}}}\label{c1} \\
  \mathrm{SINR}^{\mathrm{c}}_{n}&= \min_{k}\left( \mathrm{SINR}_{k,n}^{\mathrm{c}} \right)\label{c2} \\
  \mathrm{SINR}_{k}^{\mathrm{p}}&=\frac{\frac{\rho_{k}}{K}{\lambda}|\bg_{k,n}^{\H} \bff_{k}|^2}{{\lambda}\sum_{j\ne k}^{K}\frac{\rho_{j}}{K}|\bg_{k,n}^{\H} \bff_{j}|^2 +\sigma_{\mathrm{UE}}^{2}}\label{c3}.
  \end{align}

The achievable sum-rate is given by 
\begin{align}
 \mathrm{R}^{\!\mathrm{RS}}=\mathrm{R}^{\mathrm{c}}+\sum_{j=1}^{K}\mathrm{R}_{j}^{\mathrm{p}},\label{RSSumRate} 
\end{align}
where, similar to~\eqref{rate3}, we have $\mathrm{R}^{\mathrm{c}}\!=\!\frac{1}{T_{c}}\!\!\sum_{n=1}^{T_{c}-\tau}\!\log_{2}\!\left( 1\!+\! \mathrm{SINR}_{n}^{\mathrm{c}} \right)$  and $\mathrm{R}_{j}^{\mathrm{p}}=\frac{1}{T_{c}}\sum_{n=1}^{T_{c}-\tau}\log_{2}\left( 1+ \mathrm{SINR}_{j,n}^{\mathrm{i}} \right)$  corresponding to the common and private achievable rates, respectively. Note that $\mathrm{SINR}_{n}^{\mathrm{c}} = \displaystyle  \min_{k}\left( \mathrm{SINR}_{k,n}^{\mathrm{c}} \right)$ and $\mathrm{SINR}_{j,n}^{\mathrm{p}}$ correspond to the common and private SINRs, respectively.
\section{Deterministic Equivalent Downlink Performance Analysis with PN and Imperfect CSIT}\label{Deterministic} 
The DEs of the  SINRs for $\mathrm{NoRS}$ and $\mathrm{RS}$ are such that $\mathrm{SINR}_{k,n}-\overbar{\mathrm{SINR}}_{k,n}\xrightarrow[M \rightarrow \infty]{\mbox{a.s.}}0$\footnote{Note that $\xrightarrow[ M \rightarrow \infty]{\mbox{a.s.}}$ denotes almost sure convergence, and  $a_n\asymp b_n$ expresses the equivalence relation $a_n - b_n  \xrightarrow[ M \rightarrow \infty]{\mbox{a.s.}}  0$ with $a_n$  and $b_n$  being two infinite sequences.}, while the deterministic rate of UE $k$ is obtained by the dominated  convergence~\cite{Billingsley2008} and the continuous mapping theorem~\cite{Vaart2000} by means of~\eqref{rate3}, \eqref{RSSumRate} 
\begin{align}
R_{k}-\bar{R}_{k} \xrightarrow[ M \rightarrow \infty]{\mbox{a.s.}}0,\label{DeterministicSumrate}
\end{align}
where $\overbar{\mathrm{SINR}}_{k,n}$ and $\bar{R}_{k}$ are the corresponding DEs. However, in order to present the main results, it is priority to design the precoder for the common message.
\subsection{Precoder Design}\label{PD} 
The RS method necessitates two types of precoders multiplying the private and common messages, respectively. 
\subsubsection{Precoding of the Private Messages}
For the sake of simplicity, we design the  precoder of the private message by using RZF in terms of the channel estimate $\hat{\bG}_{n}$. Specifically, we have
\begin{align}
\bF_{n}
&= \left(\hat{\bW} \!+\! \mathrm{diag}\left(\! \hat{\bW} \!\right)\!+\!  \bZ \!+\! M \al~\! \sigma_{\mathrm{BS}}^{2} \Id_M\right)^{-1}{\hat{\bG}}\nn\\
&=  {\bSigma} {\hat{\bG}}, \label{eq:precoderRZF}
\end{align}
where ${\bSigma}\triangleq\left(\hat{\bW} \!+\!\mathrm{diag}\left(\! \hat{\bW} \!\right)\!+\!  \bZ \!+\! M \al~\! \sigma_{\mathrm{BS}}^{2} \Id_M\right)^{-1}$
with  $\hat{\bW}\triangleq\hat{\bG}\hat{\bG}^{\H}$.  Note that $\bZ \in \bbC^{M \times M}$ is an arbitrary Hermitian
nonnegative definite matrix  and $\al$ is a regularization parameter scaled by $M$, in order to converge to a constant, as $M$, $K\to \infty$. Also,   $\al$, $\bZ$ could be optimized, but this is outside the scope of this paper. 
\subsubsection{Precoding of the Common Message}
Herein, we elaborate on the design of the precoder $\bff_{c}$ of the common message in the presence of PN  by following a similar procedure to~\cite{Dai2016}. Specifically, since in the large number of antennas-regime the  different channel estimates tend to be orthogonal, we assume that $\bff_{c}$ is written as a linear sum of these channel estimates in the subspace including $\hat{\bG}$. Mathematically, this is described by 
\begin{align}
 \bff_{c}=\sum_{k}\alpha_{k} \hat{\bg}_{k}.
\end{align}

The target is the maximization of the achievable
rate of the common message $\mathrm{R}_{k,n}^{\mathrm{c}}$. This optimization problem is described by
\begin{align}\begin{split}
&\mathcal{P}_{1}~:~\max_{\bff_{c} \in \mathcal{S}}\,\min_{k} {q}_{k}|\bg_{k,n}^{\H} \bff_{c}|^2,\\
 &\mathrm{s.t.}~~~~\|\bff_{c}\|^{2}=1\label{P1} 
\end{split}
\end{align}
where $q_{k}=\frac{{\rho_{\mathrm{c}}}\lambda}{\lambda\sum_{j=1}^{K}\frac{\rho_{j}}{K}|\bg_{k,n}^{\H} \bff_{j}|^2 +\sigma_{\mathrm{UE}}^{2}}$. The optimal solution $\{\al_{k}^{*}\}$ is provided by means of the following proposition.
Note that below, we are going to use the DE of $T_{\mathrm{PN}}$, given by  Lemma~\ref{SetupLOs}.

\begin{proposition}
In the large system limit, the optimal solution of the practical problem set by $\mathcal{P}_{1}$, where PN is taken into account, is given by
\begin{align}
 \al^{*}_{k}=\frac{1}{\sqrt{M \sum_{j=1}^{K}\frac{q_{k}\frac{1}{M^{2}}\tr^{2}\hat{\bR}_{k}}{q_{j} \frac{1}{M^{2}}\tr^{2}\hat{\bR}_{j}}}},~\forall k.
\end{align}
\end{proposition}
\proof  After deriving the DEs of the equation and the constraint of the optimization problem described by $\mathcal{P}_{1}$, we lead to an optimization problem with deterministic variables. Specifically, applying~\cite[Thm. 3.7]{Hoydis2013} to \eqref{P1}, we obtain 
 \begin{align}\begin{split}
&\mathcal{P}_{2}~:~\max_{\al_{k} }\,\min_{k} {q}_{k}\frac{1}{M^{2}} |\al_{k}\tr \widetilde\bTheta_{k,n} \tr\hat{\bR}_{k}\
|^{2},\\
 &\mathrm{s.t.}~~~~\sum_{k}\al_{k}^{2}=\frac{1}{M},\label{p2} 
\end{split}
\end{align}
where  $\mathcal{P}_{2}$ includes a complex expression by means of $\widetilde\bTheta_{k,n}$.
Use of Lemma~\ref{SetupLOs} and \cite[Thm. 3.7]{Hoydis2013} transforms~\eqref{p2} to
\begin{align}
 &\mathcal{P}_{3}:\max_{\al_{k} }\min_{k}  q_{k}\al_{k}^{2} \frac{1}{M^{2}}\! \begin{cases}\!
                          \tr^{2}\hat{\bR}_{k}\!\!\!~&\mathrm{CLO~setup}\\
                             \!e^{- {\sigma_{\phi}^{2}}n}\tr^{2}\hat{\bR}_{k}~\!\!\!&\mathrm{SLOs~setup},
                       \end{cases}\\
                             &\mathrm{s.t.}~~\sum_{k}\al_{k}^{2}=\frac{1}{M}.
\end{align}
Lemma $2$ in~\cite{Xiang2014} concludes the proof by enabling us to show that the optimal solution, satisfying $\mathcal{P}_{3}$, results, if   all terms are equal. In such case, we have  $q_{k}\al_{k}^{2} \frac{1}{M^{2}}\!\!\begin{cases}\!
                          \tr^{2}\hat{\bR}_{k}=q_{j}\al_{j}^{2} \frac{1}{M^{2}}\tr^{2}\hat{\bR}_{j}&\!\!\!\!\mathrm{CLO~setup}\\
                             \!e^{- {\sigma_{\phi}^{2}}n}\tr^{2}\hat{\bR}_{k}=q_{j}\al_{j}^{2} \frac{1}{M^{2}}e^{- {\sigma_{\phi}^{2}}n}\tr^{2}\hat{\bR}_{j}&\!\!\!\!\mathrm{SLOs~setup},
                       \end{cases}$ $\forall k\ne j$
\endproof

\subsection{Achievable Deterministic  Sum-Rate with RS in the Presence of PN with Imperfect CSIT}
In this section, we conduct a DE analysis of a practical system with PN  for both the RS and the NoRS strategies. Specifically, we derive the DE of the $k$th UE in the asymptotic limit of $K, M$ for fixed ratio $\beta=K/M$.

\begin{Theorem}\label{theorem:RZF}
The downlink DEs of the SINRs of UE $k$ at time $n$ corresponding to the  private and common messages with RZF precoding in the presence of PN and imperfect CSIT, are given by
\begin{align}
\overbar{\mathrm{SINR}}_{k}^{\mathrm{p}}&=\frac{\frac{\rho_{k}}{K}\bar{\lambda}\left( \frac{\frac{1}{M}\tr \widetilde\bTheta_{k,n}{\delta}_{k}}{1+{\delta}_{k} } \right)^{2}}{{\bar{\lambda}}\sum_{j\ne k}^{K}\frac{\rho t}{K}\frac{{Q}_{jk}}{M\left(1+{\delta_{j}}\right)^{2}}+\sigma_{\mathrm{UE}}^{2}}\label{privateSINR} \\
\overbar{\mathrm{SINR}}_{k}^{c}&=\frac{{\rho_{\mathrm{c}}}\bar{\lambda}\left( \alpha_{k} \frac{1}{M}\tr\widetilde\bTheta_{k,n} \frac{1}{M}\tr \hat{\bR}_{k} \right)^2}{\bar{\lambda}{\frac{\rho t}{K}\left( \frac{\frac{1}{M}\tr \widetilde\bTheta_{k,n}{\delta}_{k}}{1+{\delta}_{k} } \right)^{2}+\sum_{j\ne k}^{K}\frac{\rho t}{K}\frac{{Q}_{jk}}{M\left(1+{\delta_{j}}\right)^{2}}+\sigma_{\mathrm{UE}}^{2}}}.\label{CommonSINR}
 \end{align}
 where 
  \begin{align}
  \bar{\lambda}&=K\left( \frac{1}{M}\sum_{k=1}^{K}\frac{{\delta}_{k}^{'}}{\left( 1+{\delta}_{k} \right)^{2}} \right)^{-1},\nn
  \end{align}
  and
  \begin{align}
\!\!{Q}_{jk}\!\asymp\! \frac{ \delta_{j}^{''}}{M}\!\!+\!\frac{\left|{\delta_{k}^{''}}\right|^{2}\delta_{k}^{''}}{M\left( 1\!+\!\delta_{j} \right)^{2}}\!-\!2\mathrm{Re}\left\{ \! \frac{\frac{1}{M}\tr\widetilde\bTheta_{k,n} {\delta}_{k}\delta_{k}^{''} }{M\left( 1\!+\!\delta_{j} \right)}\!\right\}\!.
 \label{eq:theorem2.I.mu}
\end{align}
Also, we have 
${\delta}_{k}=\frac{1}{M}\tr \hat{\bR}_{k}\bT$, $\delta_{k}^{'}=\frac{1}{M}\tr \hat{\bR}_{k}\hat{\bT}^{'}$, $\delta_{j}^{'}=\frac{1}{M}\tr \hat{\bR}_{j}\hat{\bT}^{'}$, $\delta_{k}^{''}=\frac{1}{M}\tr \hat{\bR}_{k}\hat{\bT}^{''}$, ${\bS}=\left( \kappa_{\mathrm{r}_\mathrm{UE}}^{2} \mathrm{diag}\left(\! \hat{\bR}_{k} \!\right)+\bZ \right)/M$, and $\tilde{a}=\al \sigma_{\mathrm{UE}}^{2} $
 where
\begin{itemize}
\renewcommand{\labelitemi}{$\ast$}
\item $\bT=\bT(\tilde{a})$ and ${\deltav}=[{\delta}_{1},\cdots,{\delta}_{K}]^\T={\deltav}(\tilde{a})={\ev}(\tilde{a})$ are given by \cite[Theorem 1]{Wagner2012} for ${\bS}=\bS$, $\bD_k=\tilde{\kappa}_{\mathrm{r}_\mathrm{BS}}\hat{\bR}_{k}\, \forall k \in \mathcal{K}$,
\item  $\bT^{'}=\bT^{'}(\tilde{a})$ is given by \cite[Theorem 2]{Hoydis2013} for  ${\bS}=\bS$, $\bK=\Id_M$, $\bD_k=\tilde{\kappa}_{\mathrm{r}_\mathrm{BS}}\hat{\bR}_{k}, \forall  k \in \mathcal{K}$,
\item  $\bT^{''}=\bT^{''}(\tilde{a})$ is given by \cite[Theorem 2]{Hoydis2013} for ${\bS}=\bS$, $\bK= \hat{\bR}_{k}$, $\bD_k=\tilde{\kappa}_{\mathrm{r}_\mathrm{BS}}\hat{\bR}_{k}, \forall  k \in \mathcal{K}$.
\end{itemize} 
\end{Theorem}
\proof Herein, we shall omit the proof of  Theorem~\ref{theorem:RZF}, which is provided 
in~\cite{A.Papazafeiropoulos2016} due to limited space.\endproof
\begin{remark}
Clearly, the PN affects both the numerator and the denominator of the SINRs by putting an extra penalty to the quality of the system. Its increase results in an obvious decrease of the numerator, and an increase of the denominator by means of an increase of ${Q}_{jk}$.
\end{remark}
\subsection{Power Allocation}
The optimal $t$ is rather intricate to find it  by maximizing~\eqref{RSSumRate} after calculating its first derivative. Thus, we present a suboptimal, but effective solution by following~\cite{Dai2016}. Specifically, we seek to fulfill the condition allowing RS  to outperform the conventional multi-user broadcasting. This condition is accomplished by allocating  a fraction $t$ of the
total power for the  transmission of the private messages  RS to guarantee the realization of almost the same  sum-rate as the conventional 
BC with full power. The remaining power is exploited to transmit the common message, which will enable   RS to boost the sum-rate at high-SNR. The sum-rate payoff of the RS strategy over the conventional BC (NoRS) is determined by the difference
\begin{align}
 \Delta R=\mathrm{
 R}^{\mathrm{c}}+\sum_{k=1}^{K}\left( \mathrm{R}_{k}^{\mathrm{p}}-\mathrm{R}_{k}^{\mathrm{NoRS}} \right).
\end{align}

The necessary condition and the power 
splitting ratio $t$ to achieve the outperformance are given by the following proposition.
\begin{proposition}\label{prop:inequality} 
We can write 
\begin{align}
 \mathrm{R}_{k}^{\mathrm{p}}\le 
\mathrm{R}_{k}^{\mathrm{NoRS}}.
\label{inequality} 
\end{align}
The equality holds when the power splitting ratio $t$ is given by
\begin{align}
\!\!\!t=\min\bigg\{\frac{K^{2} M }{{\bar{\lambda}}\!\sum_{j\ne k}^{K}\!\frac{\rho {Q}_{jk}}{\left(1+{\delta_{j}}\right)^{2}\!}\!},1\bigg\},\label{tau} 
\end{align}
where $\bar{m}_{k}=\tr \hat{\bR}_{k}$.
As a result, the sum-rate gain $\Delta R$ becomes
\begin{align}
 \Delta R\ge \mathrm{R}^{\mathrm{c}}-\log_2 e.
\end{align}
\end{proposition}
\proof See Appendix~\ref{proofinequality}.\endproof
\begin{remark}
According to \eqref{tau}, increasing the severity of   the PN results in less power allocated to the private messages. Interestingly, at high SNR $\rho t$ becomes independent of $\rho$, while the sum-rate
increases with the available transmit power by assigning the
remaining power $\rho-\rho t$ to the common message. On the contrary at low SNR, $t=1$, which means that the common message becomes useless. In such case, broadcasting of only  private messages takes place and RS degenerates to NoRS.  Generally, by increasing the PN, RS exhibits more its robustness. 
\end{remark}

\section{Numerical Results}\label{NumericalResults} 
This section presents the numerical illustrations of the analytical and Monte Carlo simulation results obtained for both cases of perfect and imperfect CSIT. Specifically, the bullets represent the simulation results. The black line depicts the ideal sum-rate with RZF and no common message (NoRS), i.e., perfect CSIT and hardware are assumed. The  red and cyan  as well as the blue and green lines show the sum-rate with NoRS as well as  RS  when perfect, imperfect CSIT are assumed, respectively. The discrimination between ``solid'' and ``dot'' lines  designates the  results with SLOs and CLO, respectively. 

We consider a cell  of $250~\mathrm{m}\times 250~\mathrm{m}$ with $K = 2$ UEs, where the   randomly selected UE is found at a distance of 25m from the BS. The pilot length is $B = 2$. Moreover, we assume a Rayleigh block-fading channel by  taking into consideration that the coherence time and the coherence bandwidth are $T_{c}=5~\mathrm{ms}$ and $B_{c}=100~\mathrm{KHz}$, respectively. As a result, the coherence block consists of $T = 500$ channel uses.  In each block, we assume fast fading by means of $\bw_{k} \sim \cC\cN(\b0,\Id_{M})$. Also, we set $\bR_{k}=\bm \Lambda_{k}$, i.e., we account for path-loss and shadowing, where $\Lambda_{k}$ is a $M\times M$ diagonal matrix with elements across the diagonal modeled as~\cite{Bjornson2015}
\begin{align}
 \lambda_{k}^{m}=\frac{10^{s_{k}^{m}-1.53}}{\left( d_{k}^{m} \right)^{3.76}},
\end{align}
where $d_{k}^{m}$ is the distance in meters between the  receive antenna $m$ and UE $k$, while $s_{k}^{m}\sim \cN\left(0,3.16\right)$ represents the shadowing effect. 

The  power of the uplink training symbols is  $\rho_{up}^{\mathrm{UE}}=2~\mathrm{dB}$ and the variance of thermal noise is assumed $\sigma_{\mathrm{BS}}^{2}=\sigma_{\mathrm{UE}}^{2}= -174~ \mathrm{dBm}/\mathrm{Hz}$. Moreover, the PN is simulated as a discrete Wiener process with specific increment variance.  and for the sake of simulations, we have set that the nominal values of the uplink PN equal to the downlink PN, e.g., the receive PN at the BS equals to the transmit PN at the BS.

\subsection{Impact of Hardware Impairments on NoRS/RS-Comparisons}
In the following figures, the metric under investigation is the DE sum-rate in the cases of  both NoRS  and RS strategies. The theoretical curves for the cases with imperfect  are obtained by means of Theorems~\ref{theorem:RZF}, while  the simulated curves are obtained by averaging the corresponding SINRs over $10^3$ random channel instances. Note that the curves corresponding to perfect CSIT  have been obtained, if we assume no channel estimation phase, but PN is assumed only in the downlink stage. Clearly, as can be shown from the figures,  although the DEs are   derived for $M$, $K\to \infty$ with a given ratio, the corresponding results concur with simulations for  finite values of $M$, $K$\footnote{Herein and without loss of generality, we consider the overall impact of the PN. Specifically, we add up both PN contributions coming from   BS CLO/SLOs and UE LO.}. Note that $t$ used in the simulation is  obtained  by means of  both  exhaustive search and Proposition~\ref{prop:inequality} for verification.

\begin{figure}[!h]
 \begin{center}
 \includegraphics[width=\linewidth]{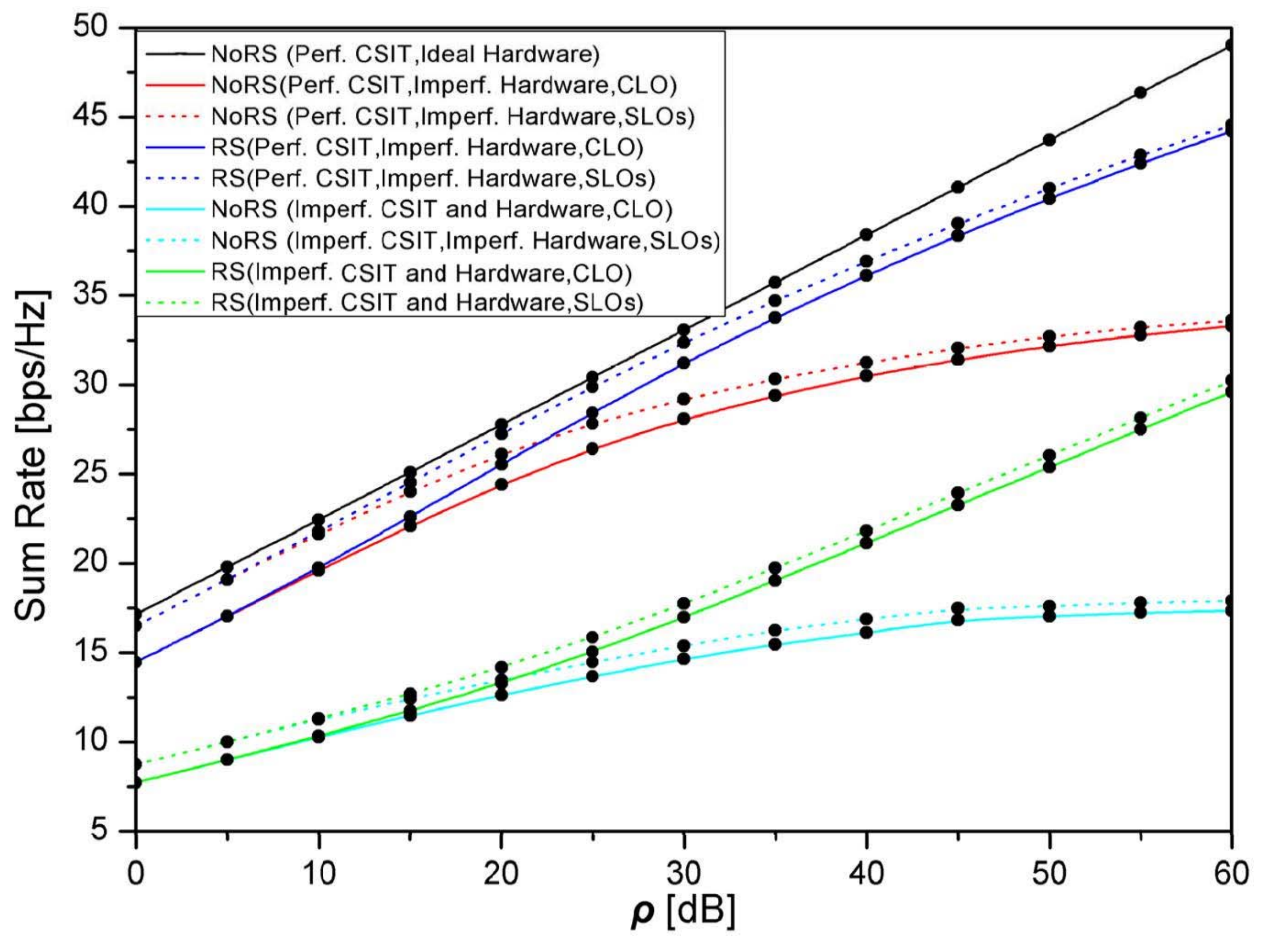}
 \caption{\footnotesize{Sum-rate versus $\rho$  ($M=100$, $K=2$, $T=500$, $\delta=10^{-4}$, $\kappa^{2}=0$).}}
 \label{M=100}
 \end{center}
 \end{figure}

Fig.~\ref{M=100} provides the comparison of the sum-rate  versus the SNR in both cases of perfect and imperfect CSIT by considering  $M=100$ and $T=500$, while   the PN is taken into account on both the uplink and the downlink.  The sum-rate with NoRS under perfect CSI and ideal hardware monotonically increases with the increase in the value of $\rho$. In the practical case where the  PN is considered, NoRS saturates after a certain value of SNR in cases of perfect (no PN and no imperfect CSIT at the uplink stage) and imperfect CSIT. Remarkably, RS proves to be robust, since the sum-rate does not saturate. Moreover, when imperfect CSIT is assumed, the degradation of the sum-rate in all cases is obvious. Notably, the setting with SLOs behaves better than the BS architecture with CLO because in such case the phase drifts are independent and in the large system limit they are averaged~\cite{Bjornson2015}. Of course, the employment of many LOs (SLOs) results in higher deployment cost. In the presence of PN, RS mitigates the multi-user interference. Thus, we have no saturation.
 
\begin{figure}[!h]
 \begin{center}
 \includegraphics[width=\linewidth]{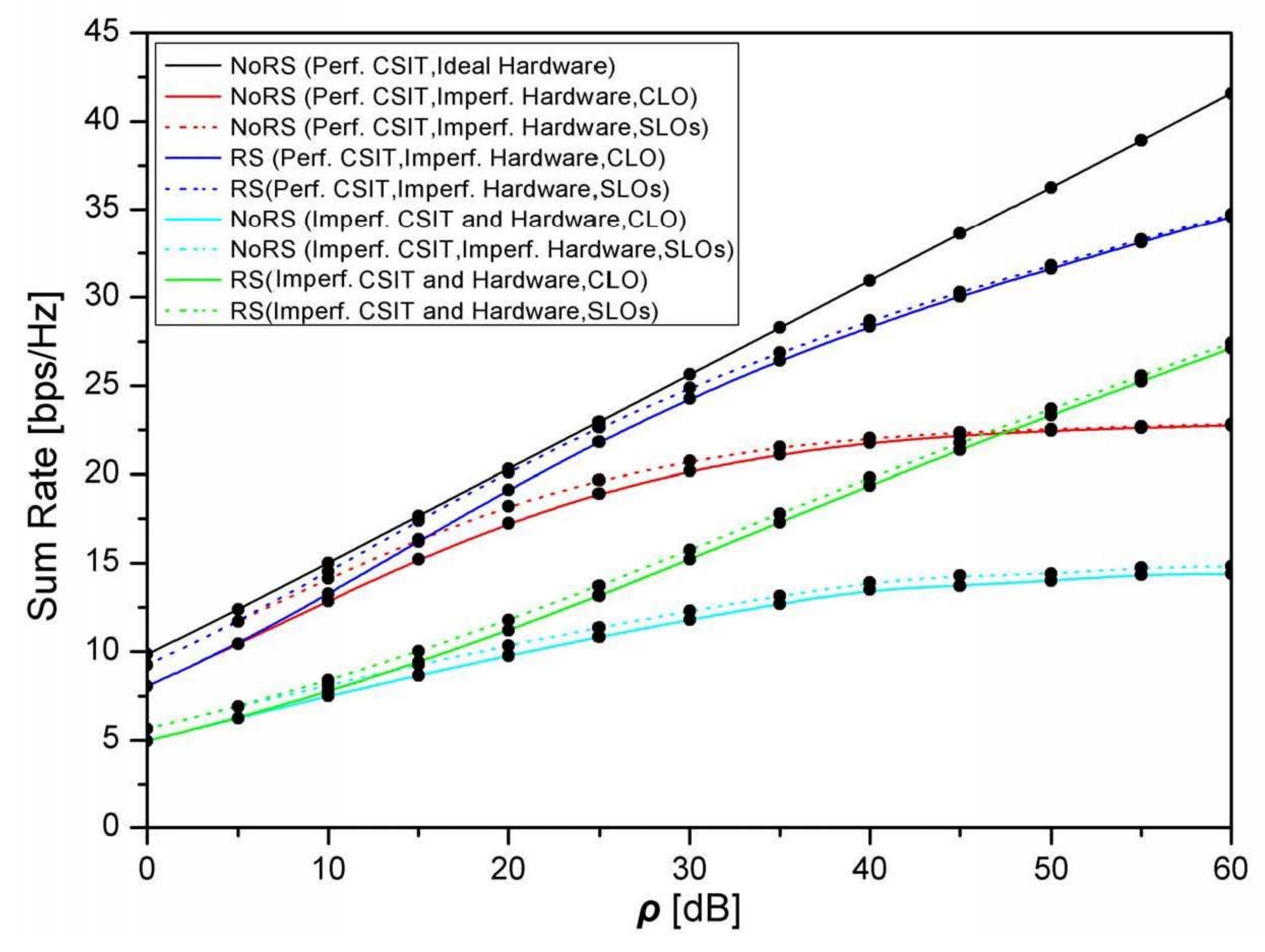}
 \caption{\footnotesize{Sum-rate versus $\rho$  ($M=20$, $K=2$, $T=500$, $\delta=10^{-4}$).}}
 \label{M=20}
 \end{center}
 \end{figure}

\begin{figure}[!h]
 \begin{center}
 \includegraphics[width=\linewidth]{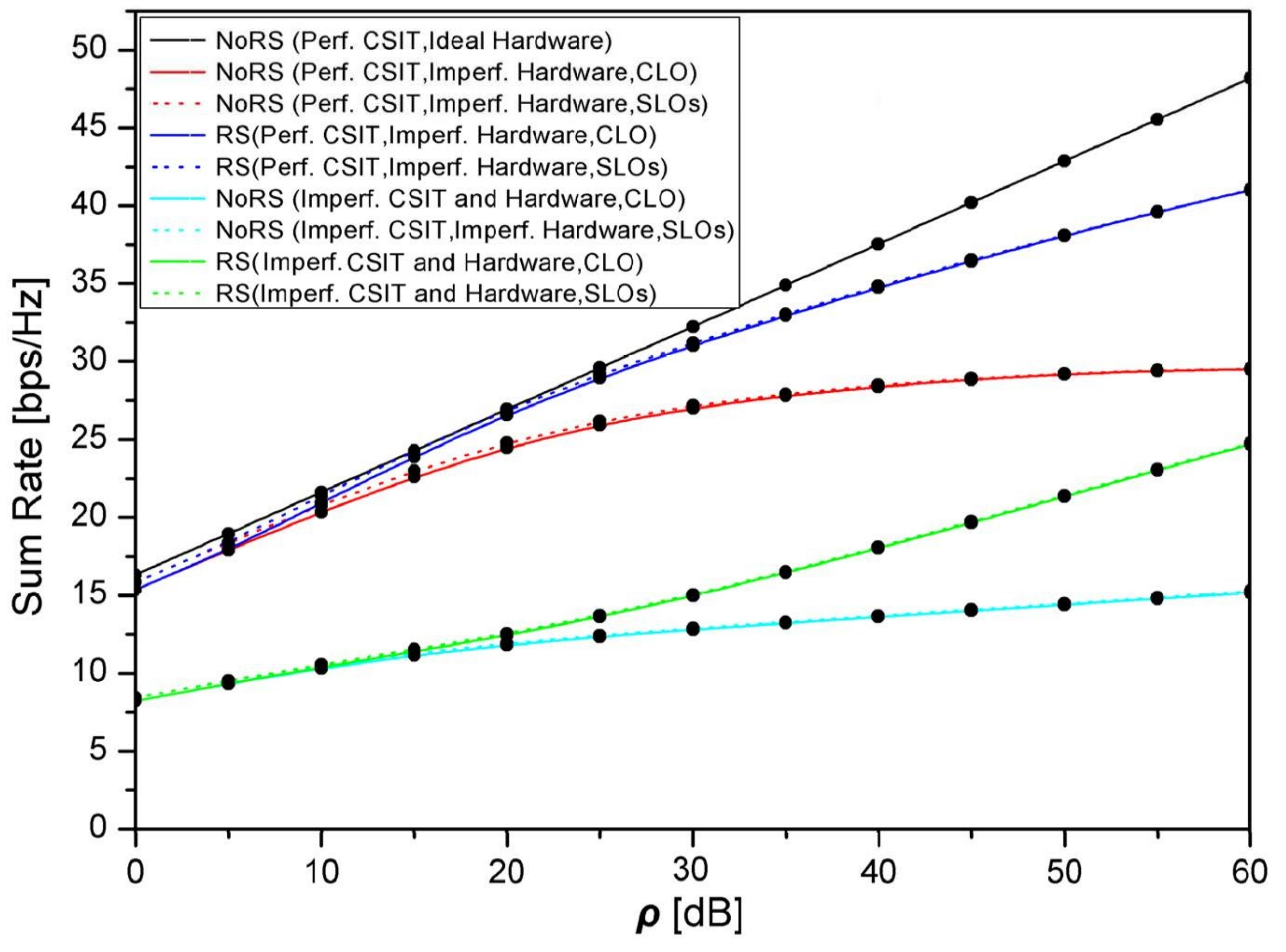}
 \caption{\footnotesize{Sum-rate versus $\rho$  ($M=100$, $K=2$, $T=100$, $\delta=10^{-4}$).}}
 \label{T=100}
 \end{center}
 \end{figure}

Fig.~\ref{M=20} presents the comparison of the  sum-rate versus the  SNR after decreasing the number of BS antennas to $M/5=20$, while Fig.~\ref{T=100} shows the impact of the coherence time $T$, when it is decreased to $T/5=100$ channel uses. In other words, in both cases we have an equivalent decrease of 1/5 of $M$ and $T$. From the former figure, we conclude a decrease of the sum-rate due to the corresponding decrease of $M$, as expected. Furthermore, the difference between the SLOs and CLO setups now is smaller because we have less independent LOs. The latter figure exposes that a decrease regarding the number of channel uses results in a small decrease of the sum-rate with the saturation in the case of NoRS taking place earlier. Also, the gap between the sum-rates with CLO and SLOs setups is smaller now.
 
\begin{figure}
\begin{center}
\subfigure[]{
\includegraphics[width=\linewidth]{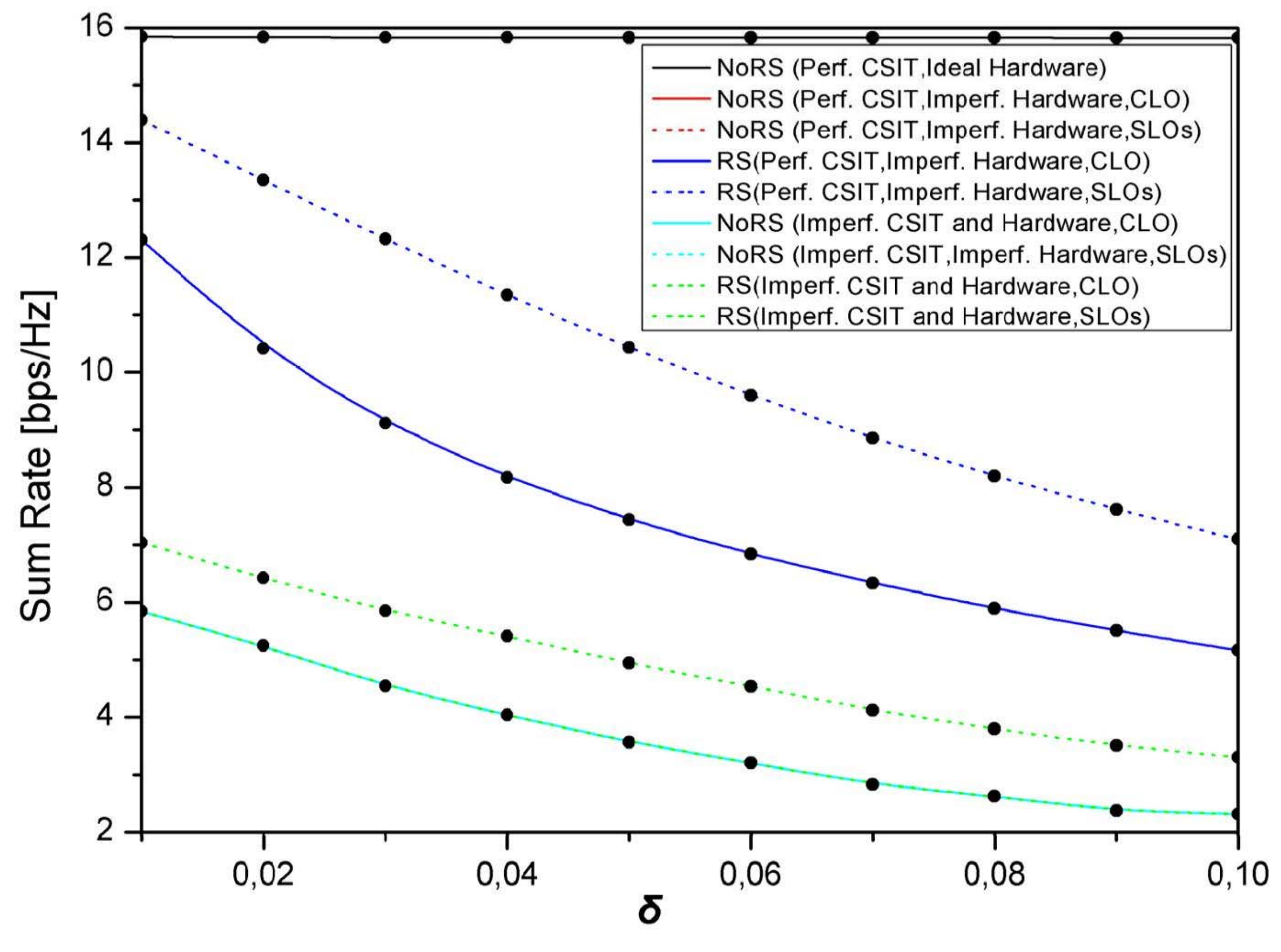}\label{25varyingdelta}
} \subfigure[]{
\includegraphics[width=\linewidth]{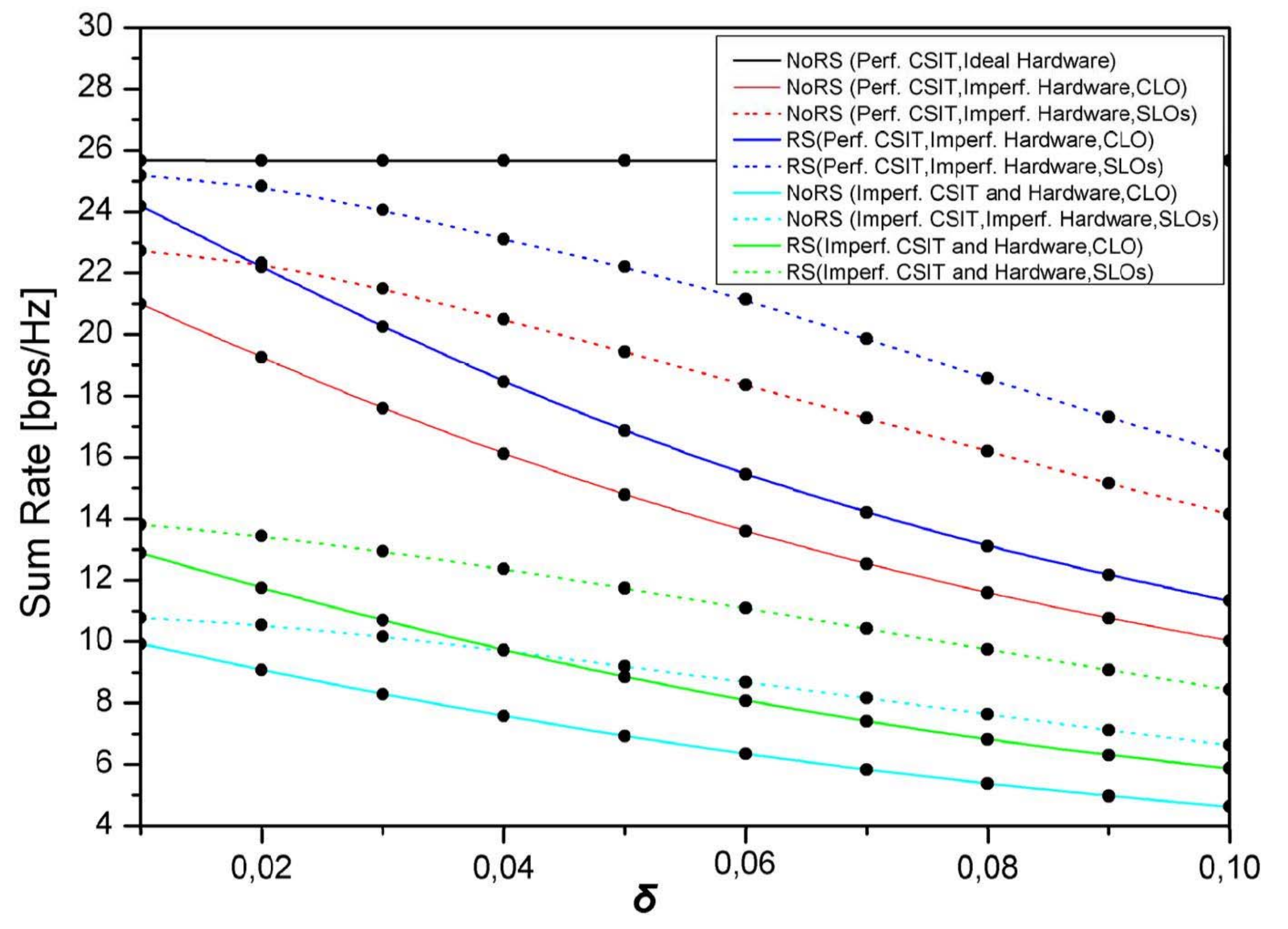}
\label{5varyingdelta}
}
\end{center}
\caption{\footnotesize{(a) Sum-rate versus $\delta$  ($M=100$, $K=2$, $T=500$, $\rho=5~\mathrm{dB}$). (b) Sum-rate versus $\delta$  ($M=100$, $K=2$, $T=500$, $\rho=25~\mathrm{dB}$).}}

\vspace{-10 pt}
\end{figure}

Figs.~\ref{25varyingdelta} and~\ref{5varyingdelta} illustrate the sum-rate versus the total PN coming from the BS CLO/SLOs and the LOs of the UEs for $\rho=5~\mathrm{dB}$ and $\rho=25~\mathrm{dB}$, respectively. According to both figures, it can be noted that the various sum-rates  decrease monotonically with $\delta$, however, in the first figure, where the SNR is small, there is no improvement coming from the implementation of RS. Hence, the NoRS lines coincide with the respective lines corresponding to the RS strategy. Obviously, the expected improvement appears in the second figure, where a higher SNR is assumed. Moreover, as $\delta$ decreases, the gap between the sum-rates corresponding to the SLOs and CLO setups narrows because the degradation coming from the accumulation of PN decreases.
\section{Conclusions}\label{Conclusions}
PN is inherent in any communication system. At the same time, the rate of multi-user systems saturates in the cases of imperfect CSIT with PN. For this reason, RS was proposed to tackle the degradation from the multi-user interference induced by imperfect CSIT. In particular, we extended the RS method to the large system regime. We pursued the DE analysis and obtained the downlink achievable rate after designing the precoders for the common and private messages. Remarkably, RS proved to be robust in the case of PN. Moreover, simulations validated the results and showed their applicability even for finite system dimensions.

\begin{appendices}
\section{Proof of Proposition~\ref{prop:inequality}}\label{proofinequality}
First, we denote  $\bar{Y}=\frac{\rho t \bar{\lambda}}{K M}{}\!\sum_{j\ne k}^{K}\!\frac{{Q}_{jk}}{\left(1+{\delta_{j}}\right)^{2}\!}$. If $\bar{Y}>1$, the private part of RS achieves the same sum-rate as the conventional multi-user BC with full power, which means that the equality in~\eqref{inequality} nearly holds. Since the common message should be decoded by all UEs, it is reasonable to   allocate less power to the common message,  as their number increases because the rate of the common decreases, i.e., the benefit of common message reduces. Thus, during the power allocation for the common message, the number of UEs should be taken into account. Moreover, following a similar rationale as in~\cite{Dai2016}, we set $\bar{Y}>K$.  We   have 
\begin{align}
\!\!\!t=\frac{K^{2} M }{{\bar{\lambda}}\!\sum_{j\ne k}^{K}\!\frac{\rho {Q}_{jk}}{\left(1+{\delta_{j}}\right)^{2}\!}\!}.\label{optimal} 
\end{align}
If we choose  $t$ as the smaller value between~\eqref{optimal} and   $1$, the inequality in~\eqref{inequality} becomes equality.
In the low-SNR regime $\rho \to 0$, \eqref{tau} gives $t=1$. In other words,
transmission of the common message is not beneficial at this regime. However, increasing the SNR, the transmission of the common message enhances the sum-rate, when the sum-rate due to only private messages tends to saturate. If we upper bound the rate loss between the private messages of the NoRS and RS, we obtain similar to~\cite{Dai2016}
\begin{align}
\sum_{j=1}^{K} \left( \mathrm{R}_{j}^{\mathrm{NoRS}}-\mathrm{R}_{j}^{\mathrm{p}} \right)
&\le \log_2 e.
\end{align}

\end{appendices}                                                                                                                                                                                                                                                                        
\bibliographystyle{IEEEtran}

\bibliography{mybib}

\end{document}